\documentclass{elsart}
\journal{Nuclear Physics B}
\begin{document}
\begin{frontmatter}
\title{Constructing a Supersymmetric Integrable System from the Hirota Method in Superspace}
\author{Debojit Sarma}
\address{Department of Physics, Cotton College, Guwahati 781001, INDIA}
\ead{debojitsarma@postmark.net}
\begin{abstract}
An  $N=1$ supersymmetric system is constructed and its integrability is shown by obtaining
three soliton solutions for it using the supersymmetric version of Hirota's direct method.
\end{abstract}
\end{frontmatter}

\section{Introduction}

The study of nonlinear evolution equations with nondispersive travelling wave 
solutions or solitons has been at the forefront for many years now.
In comparatively more recent times, the field of integrable models 
has considerably widened by the construction of integrable systems which 
are supersymmetric thus bringing into the supersymmetric regime many of the 
techniques developed in the study of bosonic or ordinary integrable systems.
Among integrable systems, the KdV equation \cite{Kort}, undoubtedly, has been 
studied most intensively. Its supersymmetric extension in $N=1$ superspace 
was formulated by Manin, Radul \cite{Manin} and Mathieu \cite{Mathieu,Mathieu1} 
in their seminal papers. Later on, Mathieu and 
co-workers \cite{Labe,Label} have found integrable extensions of the KdV equation
with $N=2$ supersymmetry and more recently other supersymmetric extensions 
of the KdV have also been reported \cite{Popo,Das}. The integrability of these 
supersymmetric equations have been proved by showing that these equations possess
one or more of the various criteria associated with integrability such as an 
infinite number of conservation laws, a bihamiltonian structure, the Lax operator 
etc. Another aspect of integrability of a nonlinear evolution is the existence of soliton 
solutions. It has been conjectured that the existence of three soliton solutions 
implies complete integrability as it appears that the three soliton
condition is intimately connected with the Painlev\'e test \cite{Ramani}.   

A number of methods have been developed to construct soliton solutions of 
nonlinear evolution equations -- the most well-known being the Inverse Scattering
Transform which was first applied by Gardner, Greene, Kruskal and Miura 
\cite{Gard} to the paradigmatic KdV equation. Another technique, 
the Hirota bilinear method \cite{Hirota}, is a powerful tool for computing
multi-soliton solutions and establishing the integrability of nonlinear
evolution equations. This method has been used by a number of workers
to extract soliton solutions of many nonlinear dynamical systems
(see for example \cite{Hiro1,Hiro2,Hiro3,Hiro4,Sats,Hiet1,Hiet2,Kake}).
The method involves first, in casting the nonlinear partial differential equation
in the bilinear form by a dependent variable transformation. This procedure is
intuitive rather than algorithmic as there is no rule for finding the correct
transformation. As a second step, which is comparatively more straightforward, 
the soliton solutions are found by substituting trial solutions for the new 
dependent variable(s) -- which is (are) usually referred to as the $\tau$ 
function(s) -- in the bilinear equations and checking for consistency. Here we
note that bilinearization of a nonlinear evolution implies the existence of one
and two soliton solutions but three and higher solitons may not exist.
It is the three soliton solution whose existence is crucial for integrability.

The Hirota formalism has also been adapted to bilinearize supersymmetric systems 
\cite{Yung,Carstea1,Carstea,Sarma2,Grama,Sarma1} and soliton solutions of 
a number of supersymmetric integrable equations have been obtained by 
the application of this technique. For instance, the $N=1$ KdV of 
Manin-Radul-Mathieu \cite{Manin,Mathieu,Mathieu1},
\begin{equation}
\partial_{t}\phi+3D^{2}(\phi D\phi)+D^{6}\phi=0
\label{1}
\end{equation}
was bilinearized \cite{Yung} via the dependent variable transformation
\begin{equation}
\phi=2D^{3}\log{\tau}
\label{2}
\end{equation}
In (\ref{1}) and (\ref{2}), $D=\partial_{\theta}+\theta\partial_{x}$ is the 
supersymmetric derivative, and $\theta$, the Grassmann odd variable. The 
transformation (\ref{2}) allows one to write the $N=1$ KdV equation in 
the bilinear form \cite{Yung}
\begin{equation}
({\bf{S}}{\bf{D}}_{t}+{\bf{S}}{\bf{D}}^{3}_{x})(\tau .\tau)=0
\label{4}
\end{equation}
where ${\bf S}$ is the supersymmetric generalization of the Hirota operator
in $N=1$ superspace and is defined by \cite{Yung}
\begin{equation}
{\bf{S}\bf{D}}^{n}_{x}(f.g)=(D_{\theta_{1}}-D_{\theta_{2}})(\partial_{x_{1}}-
\partial_{x_{2}})^{n}f(x_{1},\theta_{1})g(x_{2},\theta_{2})
{\Bigg |}_{\begin{array}{c} x_{1}=x_{2}=x .\\
\theta_{1}=\theta_{2}=\theta\end{array}}
\label{5}
\end{equation}
and ${\bf D}$ is the ordinary or bosonic Hirota operator given by
\begin{equation}
{\bf{D}}^{n}_{x}(f.g)=(\partial_{x_{1}}-\partial_{x_{2}})^{n}f(x_{1})g(x_{2})|_{x_{1}=x_{2}=x}.
\label{2c}
\end{equation} 
The soliton solutions for the $N=1$ KdV were found by the Hirota method 
\cite{Carstea1,Carstea} through the standard procedure
of substituting trial functions for the $\tau$-functions and then checking for
consistency. It was observed that the two and higher order soliton solutions could
be found only if constraints are imposed on the fermionic parameters. Subsequently, 
Carstea, Ramani and Grammaticos \cite{Grama} have shown that two and three 
soliton solutions for the $N=1$ KdV can be obtained {\em without\/} imposing 
additional fermionic constraints by dressing the fermionic part of the soliton 
solutions through the two soliton interactions and these solutions, therefore, 
constitute a more general class of soliton solutions.

In this paper, we have attempted to find if there are other supersymmetric 
nonlinear evolution equations which have general solutions of the type 
constructed in \cite{Grama}. Earlier work on the bilinearization \cite{Sarma2} 
of the $N=1$ mKdV \cite{Mathieu,Mathieu1} provides the clue to finding one such system. 
The $N=1$ mKdV 
\begin{equation}
\partial_{t}\psi+D^{6}\psi-3\psi D^{3}\psi D\psi-3(D\psi)^{2}D^{2}\psi=0
\label{6}
\end{equation}
can be converted into the following set of four bilinear equations \cite{Sarma2}
\begin{equation}
({\bf{S}}{\bf{D}}_{t}+{\bf{S}}{\bf{D}}^{3}_{x})(\tau_{1} .\tau_{1})=0 
\label{7} 
\end{equation}
\begin{equation}
({\bf{S}}{\bf{D}}_{t}+{\bf{S}}{\bf{D}}^{3}_{x})(\tau_{2} .\tau_{2})=0
\label{8}
\end{equation}
\begin{equation}
{\bf{S}}{\bf{D}}_{x}(\tau_{1} .\tau_{2})=0
\label{9}
\end{equation}
and
\begin{equation}
{\bf{D}}^{2}_{x}(\tau_{1} .\tau_{2})=0 
\label{10}
\end{equation}
after transforming the superfield $\psi$ to 
\begin{equation}
\psi=D\log\frac{\tau_{1}}{\tau_{2}}
\label{11}
\end{equation}
It was noted in \cite{Sarma2} that for the $N=1$ mKdV, in one of the
bilinear forms, namely (\ref{10}), the supersymmetric Hirota operator
{\bf S} does not appear. As a result, two and three soliton solutions without constraints 
on the fermionic parameters could {\em not\/} be obtained for it. However, 
if a supersymmetric system which had bilinear forms (\ref{7}), (\ref{8}) 
and (\ref{9}) only could be constructed, one would have constraint-free
soliton solutions of the type obtained in \cite{Grama}.

In the next section, we construct a supersymmetric system in $N=1$ 
superspace that can be transformed into the bilinear forms (\ref{7}), (\ref{8}) 
and (\ref{9}). Moreover, the supersymmetric system has the KdV as its bosonic 
limit and is therefore a supersymmetric extension of it. In Section 3, one, 
two and three soliton solutions for it are written down.  These soliton solutions
are of the type found in \cite{Grama} {\it i.e.} free from constraints on the 
fermionic parameters. Section 4 is the concluding one.

\section{An $N=1$ Extension of the KdV}

Consider the following pair of supersymmetric equations in $N=1$ superspace:
\begin{equation}
\partial_{t}\phi+\partial_{x}^{3}\phi-6\partial_{x}[(D\phi)\phi\ D^{-1}\phi]
-2\partial_{x}\phi^{3}+3\partial_{x}(\psi D\phi)=0
\label{12}
\end{equation}
\begin{equation}
\partial_{t}\psi+\partial_{x}^{3}\psi-6\partial_{x}[(D\psi)\phi\ D^{-1}\phi]
-3\partial_{x}(\psi D\psi)+3\partial_{x}[(\partial_{x}\phi) D\phi]=0
\label{13}
\end{equation}
where $D^{-1}=\theta+\partial_{\theta}\partial_{x}^{-1}$ is the formal inverse of the 
operator $D$ {\it i.e.} $DD^{-1}=D^{-1}D=1$. Clearly, (\ref{12},\ref{13}) are 
nonlocal, coupled, partial differential equations in $N=1$ superspace for the 
superfields $\phi$ and $\psi$ which have conformal weights $1$ and $3 \over 2$ 
respectively. If these superfields are written in terms of their bosonic and 
fermionic components
\begin{equation}
\phi=\phi_{b}+\theta\phi_{f}
\label{14}
\end{equation}
\begin{equation}
\psi=\psi_{f}+\theta\psi_{b}
\label{15}
\end{equation}
then (\ref{12}) and (\ref{13}) can be split up into four equations in terms of 
the component fields. For (\ref{12}), they are
\begin{eqnarray}
&&\partial_{t}\phi_{b}+\partial_{x}^{3}\phi_{b}-6\partial_{x}(\phi_{f}\phi_{b}
\partial^{-1}_{x}\phi_{f})-2\partial_{x}\phi^{3}_{b}+3\partial_{x}
(\psi_{f}\phi_{f})=0
\label{16} \\
&&\partial_{t}\phi_{f}+\partial_{x}^{3}\phi_{f}-6\partial_{x}[(\partial_{x}\phi_{b})
\phi_{b}\partial^{-1}_{x}\phi_{f}]+3\partial_{x}(\psi_{b}\phi_{f})-3\partial_{x}
(\psi_{f}\partial_{x}\phi_{b})=0
\label{17} 
\end{eqnarray}
and the component equations for (\ref{13}) are
\begin{eqnarray}
&&\partial_{t}\psi_{f}+\partial_{x}^{3}\psi_{f}-6\partial_{x}(\psi_{b}\phi_{b}
\partial^{-1}_{x}\phi_{f})-3\partial_{x}(\psi_{f}\psi_{b})+3\partial_{x}
(\phi_{f}\partial_{x}\phi_{b})=0
\label{18} \\
&&\partial_{t}\psi_{b}+\partial^{3}_{x}\psi_{b}-6\partial_{x}(\psi_{b}\phi^{2}_{b})
-6\psi_{b}\partial_{x}\psi_{b}+3\partial_{x}(\psi_{f}\partial_{x}\psi_{f})
+3\partial_{x}(\partial_{x}\phi_{b})^{2} \nonumber \\
&&-6\partial_{x}(\psi_{b}\phi_{f}\partial^{-1}_{x}\phi_{f})
-6\partial_{x}[(\partial_{x}\psi_{f})\phi_{b}\partial^{-1}_{x}\phi_{f}]
-3\partial_{x}(\phi_{f}\partial_{x}\phi_{f})=0
\label{19}
\end{eqnarray}
Equations (\ref{16}), (\ref{17}), (\ref{18}) and (\ref{19})
are invariant under the $N=1$ supersymmetric transformations
\begin{equation}
\delta\phi_{b}=\eta\phi_{f} \;\;\;\;\;\;\;
\delta\phi_{f}=\eta\partial_{x}\phi_{b} 
\label{20a}
\end{equation}
\begin{equation}
\delta\psi_{b}=\eta\partial_{x}\psi_{f} \;\;\;\;\;\;\; 
\delta\psi_{f}=\eta\psi_{b}
\label{20b}
\end{equation} 
and therefore constitute an $N=1$ supersymmetric system. Moreover, on setting
the fields $\phi_{b}$, $\phi_{f}$ and $\psi_{f}$ to zero we obtain the KdV
equation
\begin{equation}
\partial_{t}\psi_{b}+\partial^{3}_{x}\psi_{b}-6\psi_{b}\partial_{x}\psi_{b}=0
\label{21}
\end{equation}
This clearly indicates that equations (\ref{12},\ref{13}) may be considered 
a supersymmetric extension of the KdV equation. Note that, as in case of the 
$N=2$ KdV \cite{Labe}, the bosonic core of the equations (\ref{12}) and 
(\ref{13}) also contain the mKdV equation 
$\partial_{t}\phi_{b}+\partial_{x}^{3}\phi_{b}-2\partial_{x}\phi^{3}_{b}=0$.

In order to write the supersymmetric equations (\ref{12}) and (\ref{13}) in 
bilinear form, consider the following transformations for the superfields 
$\phi$ and $\psi$:
\begin{equation}
\phi=D^{2}\log {{\tau_{1}}\over {\tau_{2}}}
\label{22}
\end{equation}
and
\begin{equation}
\psi=D^{3}\log (\tau_{1} \tau_{2})
\label{23}
\end{equation}
where $\tau_{1}$ and $\tau_{2}$  are independent, bosonic superfields. Substituting 
for the fields $\phi$ and $\psi$ in the equations (\ref{12}) and (\ref{13}), 
we find that they can be cast into a form involving only bilinears. For instance,
equation (\ref{12}) after substitution acquires the form
\begin{eqnarray}
&&\partial_{t}D^{2}\log\frac{\tau_{1}}{\tau_{2}}+D^{8}\log\frac{\tau_{1}}{\tau_{2}}
-6D^{2}\left[D^{3}\log\frac{\tau_{1}}{\tau_{2}}\; D^{2}\log\frac{\tau_{1}}{\tau_{2}}
\; D\log\frac{\tau_{1}}{\tau_{2}}\right] \nonumber \\
&&-2D^{2}\left[D^{2}\log\frac{\tau_{1}}{\tau_{2}}\right]^{3}
+3D^{2}\left[D^{3}\log (\tau_{1}\tau_{2})\; D^{3}\log\frac{\tau_{1}}{\tau_{2}}\right]=0
\label{22a}
\end{eqnarray}
Adding and subtracting the term 
$6D^{4}\log\frac{\tau_{1}}{\tau_{2}}\; D^{3}\log (\tau_{1}\tau_{2})$ to 
the left hand side of (\ref{22a}) after operating on it by $D^{-1}$ we get,
\begin{eqnarray} 
&&\left[2\partial_{t}D\log\tau_{1}+2D^{7}\log\tau_{1}+12D^{4}\log\tau_{1}\; D^{3}\log\tau_{1} 
\right]\nonumber \\
&&-\left[2\partial_{t}D\log\tau_{2}+2D^{7}\log\tau_{2}+
12D^{4}\log\tau_{2}\; D^{3}\log\tau_{2}\right] \nonumber\\
&&-12\left(D^{4}\log\tau_{1}-D^{4}\log\tau_{2}\right)\left[D^{3}\log (\tau_{1}\tau_{2})
+D^{2}\log\frac{\tau_{1}}{\tau_{2}}\; D\log\frac{\tau_{1}}{\tau_{2}}\right]=0
\label{22b}
\end{eqnarray}
Writing the above equation using the Hirota operators, we find that equation 
(\ref{12}) is finally transformed into
\begin{eqnarray}
&&\frac{({\bf{S}}{\bf{D}}_{t}+{\bf{S}}{\bf{D}}^{3}_{x})(\tau_{1}.\tau_{1})}{\tau_{1}^{2}}
-\frac{({\bf{S}}{\bf{D}}_{t}+{\bf{S}}{\bf{D}}^{3}_{x})(\tau_{2}.\tau_{2})}{\tau_{2}^{2}}
\nonumber \\
-&&6\left[\frac{{\bf{D}}^{2}_{x}(\tau_{1}.\tau_{1})}{\tau_{1}^{2}}
-\frac{{\bf{D}}^{2}_{x}(\tau_{2}.\tau_{2})}{\tau_{2}^{2}}\right]
\frac{{\bf{S}}{\bf{D}}_{x}(\tau_{1}.\tau_{2})}{\tau_{1}\tau_{2}}=0
\label{24}
\end{eqnarray}
An almost identical calculation transforms equation (\ref{13}) into
\begin{eqnarray}
&&\frac{({\bf{S}}{\bf{D}}_{t}+{\bf{S}}{\bf{D}}^{3}_{x})(\tau_{1}.\tau_{1})}{\tau_{1}^{2}}
+\frac{({\bf{S}}{\bf{D}}_{t}+{\bf{S}}{\bf{D}}^{3}_{x})(\tau_{2}.\tau_{2})}{\tau_{2}^{2}}
\nonumber \\-&&6\frac{{\bf{S}}{\bf{D}}_{x}(\tau_{1}.\tau_{2})}{\tau_{1}\tau_{2}}
\left[\frac{{\bf{D}}^{2}_{x}(\tau_{1}.\tau_{1})}{\tau_{1}^{2}}
+\frac{{\bf{D}}^{2}_{x}(\tau_{2}.\tau_{2})}{\tau_{2}^{2}}\right]=0
\label{24a}
\end{eqnarray}
From (\ref{24}) and (\ref{24a}), it follows that the supersymmetric 
system (\ref{12},\ref{13}) has the bilinear forms given by (\ref{7}), (\ref{8}) and
(\ref{9}), that is only those equations which involve the super Hirota operator
${\bf S}$. Clearly, unlike the $N=1$ mKdV equation, this supersymmetric system does not 
have any bilinear equation involving only bosonic Hirota operators.

\section{Soliton Solutions}

For one soliton solutions, consider the following forms for the $\tau$ functions:
\begin{equation}
\tau_{1}=1+e^{\eta +\theta\zeta}
\label{25}
\end{equation}       
and
\begin{equation}
\tau_{2}=1-e^{\eta +\theta\zeta}
\label{26}
\end{equation}   
where $\eta=kx+\omega t$. Here $k$ and $\omega$ are the bosonic parameters 
while $\zeta$ is the Grassmann odd parameter. These are solutions of (\ref{7}) 
and (\ref{8}) provided $\omega=-k^{3}$ which is the KdV dispersion relation 
and in addition, the $\tau$ functions (\ref{25},\ref{26}) satisfy (\ref{9})
and are therefore appropriate $\tau$ functions for the set bilinear equations 
(\ref{7},\ref{8},\ref{9}) 
representing (\ref{12},\ref{13}). Throughout this section, we employ the 
notation similar to that used in \cite{Grama} for the $\tau$ functions of the 
$N=1$ KdV.

In obtaining the two soliton solution for the $N=1$ mKdV equation \cite{Sarma2},
it was found that the solutions exist provided the fermionic parameters $\zeta_{1}$
and $\zeta_{2}$ are related via the constraint relation 
$k_{1}\zeta_{2}=k_{2}\zeta_{1}$ 
and identical relations characterize the three soliton solutions also. For the 
supersymmetric equations (\ref{12},\ref{13}) however, by an appropriate choice 
of $\tau$ functions, general solutions can be obtained. Explicitly, the $\tau$ 
functions are  
\begin{equation}
\tau_{1}=1+e^{\eta_{1}+\theta\zeta_{1}}+e^{\eta_{2}+\theta\zeta_{2}}+
A_{12}\left[1+2\frac{\zeta_{1}\zeta_{2}}{k_{2}-k_{1}}\right]
e^{\eta_{1}+\eta_{2}+\theta(\alpha_{12}\zeta_{1}+\alpha_{21}\zeta_{2})}
\label{27}
\end{equation} 
and
\begin{equation}
\tau_{2}=1-e^{\eta_{1}+\theta\zeta_{1}}-e^{\eta_{2}+\theta\zeta_{2}}+
A_{12}\left[1+2\frac{\zeta_{1}\zeta_{2}}{k_{2}-k_{1}}\right]
e^{\eta_{1}+\eta_{2}+\theta(\alpha_{12}\zeta_{1}+\alpha_{21}\zeta_{2})}
\label{28}
\end{equation}
where $\eta_{i}=k_{i}x+\omega_{i}t$, $(i=1,2)$ and 
\begin{equation}
A_{12}=\left({{k_{1}-k_{2}}\over {k_{1}+k_{2}}}\right)^{2}
\label{28a}
\end{equation}
is the bosonic interaction term. The $\alpha_{ij}\;'s$  which are responsible
for the dressing of the fermionic part through the two soliton interactions \cite{Grama}
are given by
\begin{equation} 
\alpha_{ij}={{k_{i}+k_{j}}\over {k_{i}-k_{j}}} \;\;\;\;\;(i,j=1,2).
\label{28b}
\end{equation}
Substituting (\ref{27}) in (\ref{7}) and (\ref{28}) in (\ref{8}) yields the 
dispersion relations $\omega_{i}=-k^{3}_{i}$ with $i=1,2$. These $\tau$ 
functions also  satisfy (\ref{9}) and are therefore the appropriate $\tau$
functions for the system considered here.
 
The $\tau$ functions for the three soliton solutions of the system (\ref{12},\ref{13})
have the form 
\begin{eqnarray}
&&\tau_{1}=1+e^{\eta_{1}+\theta\zeta_{1}}+e^{\eta_{2}+\theta\zeta_{2}}
+e^{\eta_{3}+\theta\zeta_{3}} \nonumber \\ 
&&+A_{12}\left[1+2\frac{\zeta_{1}\zeta_{2}}{k_{2}-k_{1}}\right]
e^{\eta_{1}+\eta_{2}+\theta(\alpha_{12}\zeta_{1}+\alpha_{21}\zeta_{2})} \nonumber \\
&&+A_{13}\left[1+2\frac{\zeta_{1}\zeta_{3}}{k_{3}-k_{1}}\right]
e^{\eta_{1}+\eta_{3}+\theta(\alpha_{13}\zeta_{1}+\alpha_{31}\zeta_{3})} \nonumber \\
&&+A_{23}\left[1+2\frac{\zeta_{2}\zeta_{3}}{k_{3}-k_{2}}\right]
e^{\eta_{2}+\eta_{3}+\theta(\alpha_{23}\zeta_{2}+\alpha_{32}\zeta_{3})} \nonumber \\
&&+A_{12}A_{13}A_{23}\left[1+2\frac{(\alpha_{13}\zeta_{1})(\alpha_{23}\zeta_{2})}{k_{2}-k_{1}}
\right]\left[1+2\frac{(\alpha_{12}\zeta_{1})(\alpha_{32}\zeta_{3})}{k_{3}-k_{1}}
\right] \nonumber \\
&&\times\left[1+2\frac{(\alpha_{21}\zeta_{2})(\alpha_{31}\zeta_{3})}{k_{3}-k_{2}}
\right]e^{\eta_{1}+\eta_{2}+\eta_{3}+\theta(\alpha_{12}\alpha_{13}\zeta_{1}
+\alpha_{21}\alpha_{23}\zeta_{2}+\alpha_{31}\alpha_{32}\zeta_{3})}
\label{29}
\end{eqnarray}
and
\begin{eqnarray}
&&\tau_{1}=1-e^{\eta_{1}+\theta\zeta_{1}}-e^{\eta_{2}+\theta\zeta_{2}}
-e^{\eta_{3}+\theta\zeta_{3}} \nonumber \\
&&+A_{12}\left[1+2\frac{\zeta_{1}\zeta_{2}}{k_{2}-k_{1}}\right]
e^{\eta_{1}+\eta_{2}+\theta(\alpha_{12}\zeta_{1}+\alpha_{21}\zeta_{2})} \nonumber \\
&&+A_{13}\left[1+2\frac{\zeta_{1}\zeta_{3}}{k_{3}-k_{1}}\right]
e^{\eta_{1}+\eta_{3}+\theta(\alpha_{13}\zeta_{1}+\alpha_{31}\zeta_{3})} \nonumber \\
&&+A_{23}\left[1+2\frac{\zeta_{2}\zeta_{3}}{k_{3}-k_{2}}\right]
e^{\eta_{2}+\eta_{3}+\theta(\alpha_{23}\zeta_{2}+\alpha_{32}\zeta_{3})} \nonumber \\
&&-A_{12}A_{13}A_{23}\left[1+2\frac{(\alpha_{13}\zeta_{1})(\alpha_{23}\zeta_{2})}{k_{2}-k_{1}}
\right]\left[1+2\frac{(\alpha_{12}\zeta_{1})(\alpha_{32}\zeta_{3})}{k_{3}-k_{1}}
\right] \nonumber \\
&&\times\left[1+2\frac{(\alpha_{21}\zeta_{2})(\alpha_{31}\zeta_{3})}{k_{3}-k_{2}}
\right]e^{\eta_{1}+\eta_{2}+\eta_{3}+\theta(\alpha_{12}\alpha_{13}\zeta_{1}
+\alpha_{21}\alpha_{23}\zeta_{2}+\alpha_{31}\alpha_{32}\zeta_{3})}
\label{29a}
\end{eqnarray}
Here the bosonic interaction terms $A_{ij}$ are given by
\begin{equation}
A_{ij}=\left({{k_{i}-k_{j}}\over {k_{i}+k_{j}}}\right)^{2} \;\;\;\;\; (i,j=1,2,3)
\label{29c}
\end{equation}
and the $\alpha_{ij}\;'s$ are obtained from (\ref{28b}) with $i,j=1,2,3$.
We have verified that these $\tau$ functions satisfy the 
bilinear equations (\ref{7}), (\ref{8}) and (\ref{9}) and hence three soliton 
solutions exist for the supersymmetric system given by  (\ref{12},\ref{13}).
Also note that function $\tau_{1}$ for the one, two and three soliton
discussed above are exactly those found by Carstea-Ramani-Grammaticos \cite{Grama}
while the corresponding $\tau_{2}\;'s$ differ by the sign of some of the terms.
The relation between structures of the $\tau$ functions for the $N=1$ KdV \cite{Grama}
and those written down here parallels the relations between the forms of the $\tau$ 
functions for the bosonic KdV \cite{Hirota} and the mKdV \cite{Hiro1}. 

Thus the supersymmetric evolution equations (\ref{12},\ref{13}) have at least three
soliton solutions which are free from fermionic constraints. Since the existence of 
three soliton solution implies integrability, (\ref{12},\ref{13}) constitute an $N=1$ 
supersymmetric integrable extension of the KdV equation. 

It is also relevant to mention that for extensions of the KdV with a single
additional fermionic field, there are only two which are know to be integrable -- the super KdV of 
Manin-Radul-Mathieu \cite{Manin,Mathieu,Mathieu1} which is invariant under $N=1$ supersymmetric 
transformations and the one proposed
by Kupershmidt \cite{Kuper} which is not. Among $N=2$ extensions, i.e. those
with two fermionic and an additional bosonic field and also invariant under $N=2$ supersymmetry,
five are known to be integrable. They are the three $N=2$ super KdV equations which are integrable 
for a free parameter $a$ having values $2,-1,4$ \cite{Labe,Label}, the odd $N=2$ super KdV \cite{Popo}
and the super KdV-B \cite{Das}. Although system consisting of equations (\ref{12},\ref{13}) discussed in 
this paper is a superextension of the KdV with two fermions and an extra bosonic field,
it is however not invariant under $N=2$ supersymmetry but is an $N=1$ supersymmetric, integrable extension of
the KdV. A distinctive feature of this system is its nonlocality which is not observed in the 
integrable extensions of the KdV that are already classified.

\section{Conclusion}

In this paper, we have constructed a pair of coupled $N=1$ supersymmetric 
evolution equations whose bosonic limit contains the KdV equation. This 
system was constructed in such a way that it can be transformed into a set of 
Hirota bilinear equations which possess at least three soliton solutions free from 
constraints.  Since three soliton solutions exist for the system, the system may be 
considered an integrable, $N=1$ supersymmetric extension of the KdV equation.

\ack{The author is grateful to Prof. P. Mathieu for expressing his view on integrability
in response to the author's query on the subject.}

\end{document}